\documentclass{JINST}
\usepackage{lineno}
\usepackage[percent]{overpic}
\usepackage{amsmath}
\usepackage{amssymb}
\usepackage{graphics}
\newcommand{\DBD}{0$\nu$DBD}

\newcommand{\TEO}{$\mathrm{TeO}_2$}
\newcommand{\TEHT}{$^{130}\mathrm{Te}$}

\newcommand{\Cuoricino}{CUORICINO}
\newcommand{\Cuore}{CUORE}
\newcommand{\LNGS}{LNGS}

\providecommand*{\un}[1]{\ensuremath{\mathrm{~#1}}}

\title{Noise correlation and decorrelation in arrays of bolometric detectors}
\author{
C.~Mancini-Terracciano$^{a,1}$ and M.~Vignati$^{a}$\thanks{Corresponding author.}\\
\llap{$^a$}Dipartimento di Fisica, Sapienza Universit\`a di Roma and Sezione INFN di Roma,\\ Piazzale~A.~Moro~2, Roma I-00185, Italy\\
\llap{$^1$}now at 
Dipartimento di Fisica, Universit\`a degli Studi Roma Tre, \\ Via della Vasca Navale 84, Roma I-00146, Italy\\
E-mail:\email{marco.vignati@roma1.infn.it}
}

\abstract{
Bolometers are phonon mediated detectors used in particle physics
experiments to search for rare processes, such as neutrinoless double beta
decay and dark matter interactions.
They feature an excellent energy resolution,
which is a few keV over an energy range extending from a few keV up to several
MeV. Nevertheless the resolution can be limited by the noise induced by
vibrations of the mechanical apparatus. In arrays of bolometers part of this noise is correlated
among different detectors and can be removed using a multichannel decorrelation algorithm. 
In this paper we present a decorrelation method and its application to data from the
\Cuoricino\ experiment, an array of 62 \TEO\ bolometers.
}

\keywords{Bolometers, Thermal noise, Digital signal processing}

\begin{document}

\section{Introduction} 

Bolometers are detectors in which the energy from particle interactions
is converted to thermal energy and measured via their rise in temperature. They
provide excellent energy resolution, though their response is slow
compared to conventional detectors.  These features make them a suitable
choice for experiments searching for rare processes, such as neutrinoless
double beta decay (\DBD) and dark matter (DM) interactions.

The \Cuore\ experiment will search for \DBD\ of \TEHT~\cite{Ardito:2005ar,
ACryo} using an array of 988 \TEO\ bolometers of 750\un{g} each. Operated
at a temperature of about 10\un{mK}, these detectors feature an
energy resolution of a few keV over their energy range, extending
from a few keV up to several MeV.  At low energies, below 100 keV, 
the resolution is of the order of 1 keV FWHM, while at 2528\un{keV}, the
\DBD\ energy, is about 5\un{keV\,FWHM}; this,
together with the low background and the high mass of the experiment,
determines the sensitivity to the \DBD.  \Cuore\ could also search for
DM interactions, provided that the energy threshold is of
few keV~\cite{Didomizio:2010ph}. An experiment made of 62 bolometers,
\Cuoricino~\cite{Andreotti:2010vj}, was operated at Laboratori Nazionali
del Gran Sasso (\LNGS) in Italy between 2003 and 2008, and proved the
feasibility of the \TEO\ bolometric technique. \Cuore\ is currently under
construction, and will start taking data in three years.

Vibrations of the cryogenic apparatus (keeping the system at
low temperatures) induce a visible noise, which limits the
energy resolution at low energies~\cite{Bellini:2010iw} and the energy
threshold~\cite{Didomizio:2010ph}.  Since  all bolometers are held by the same
structure, part of the  vibrational noise is expected to be correlated.

In this paper we present a method to estimate the correlated
noise among different bolometers, and a method to remove it.
The application to data from \Cuoricino\ shows that the
correlated noise is visible and that it can be efficiently removed.

The \Cuore\ detector structure will be different from the \Cuoricino\ one,
and thus a different vibrational noise is expected. If it will be unfortunately large, 
the algorithms developed in this work could be used to improve the performances of \Cuore\ in both \DBD\ and DM searches.

\section{Experimental setup}\label{sec:experimental setup}

\Cuoricino\ and \Cuore\ bolometers are composed of two main parts,
a \TEO\ crystal and a neutron transmutation doped Germanium (NTD-Ge)
thermistor ~\cite{wang,Itoh}. The crystal is cube-shaped (5x5x5~cm$^{3}$)
and held by Teflon supports in copper frames. The frames are connected
to the mixing chamber of a dilution refrigerator, which keeps the
system at a temperature of about $10\un{mK}$. The thermistor is glued
to the crystal and acts as thermometer. When energy is released in
the crystal, its temperature increases and changes the resistance of the thermistor.
To read out the signal,
the thermistor is biased with constant current, which is provided by a
voltage generator and a load resistor in series with the thermistor.
The resistance of the thermistor varies in time with the temperature,
and the voltage across it is the bolometer signal.  The value of the
load resistor is chosen to be much higher than the thermistor, so that
the voltage across the thermistor is proportional to its resistance.
The wires that connect the thermistor to the electronics introduce a
non-negligible parasitic capacitance. 

\begin{figure}[htbp]
\centering
\begin{minipage}{0.48\textwidth}
\includegraphics[clip=true,width=1\textwidth]{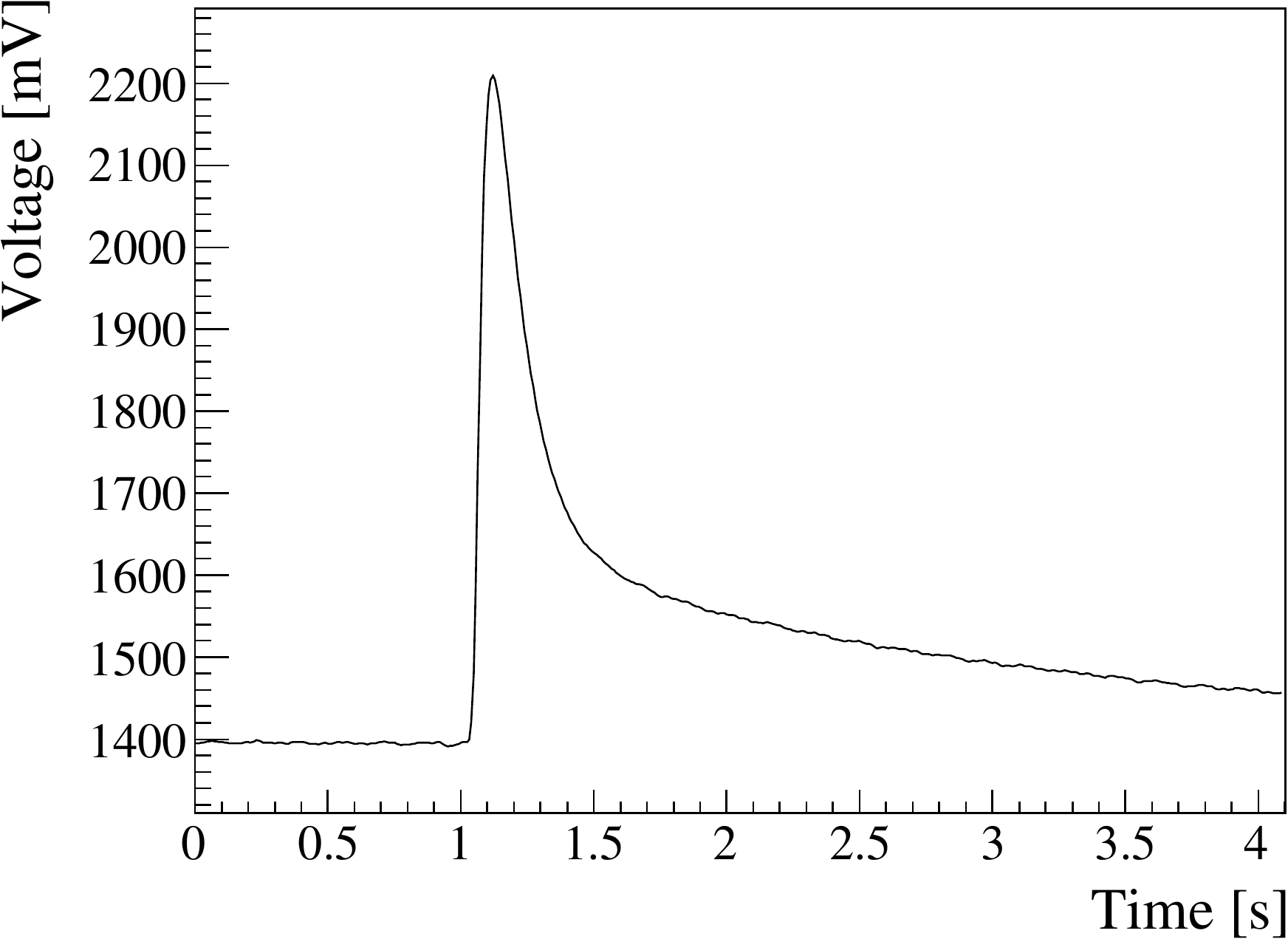}  
\end{minipage}
\hfill
\begin{minipage}{0.48\textwidth}
\includegraphics[clip=true,width=1\textwidth]{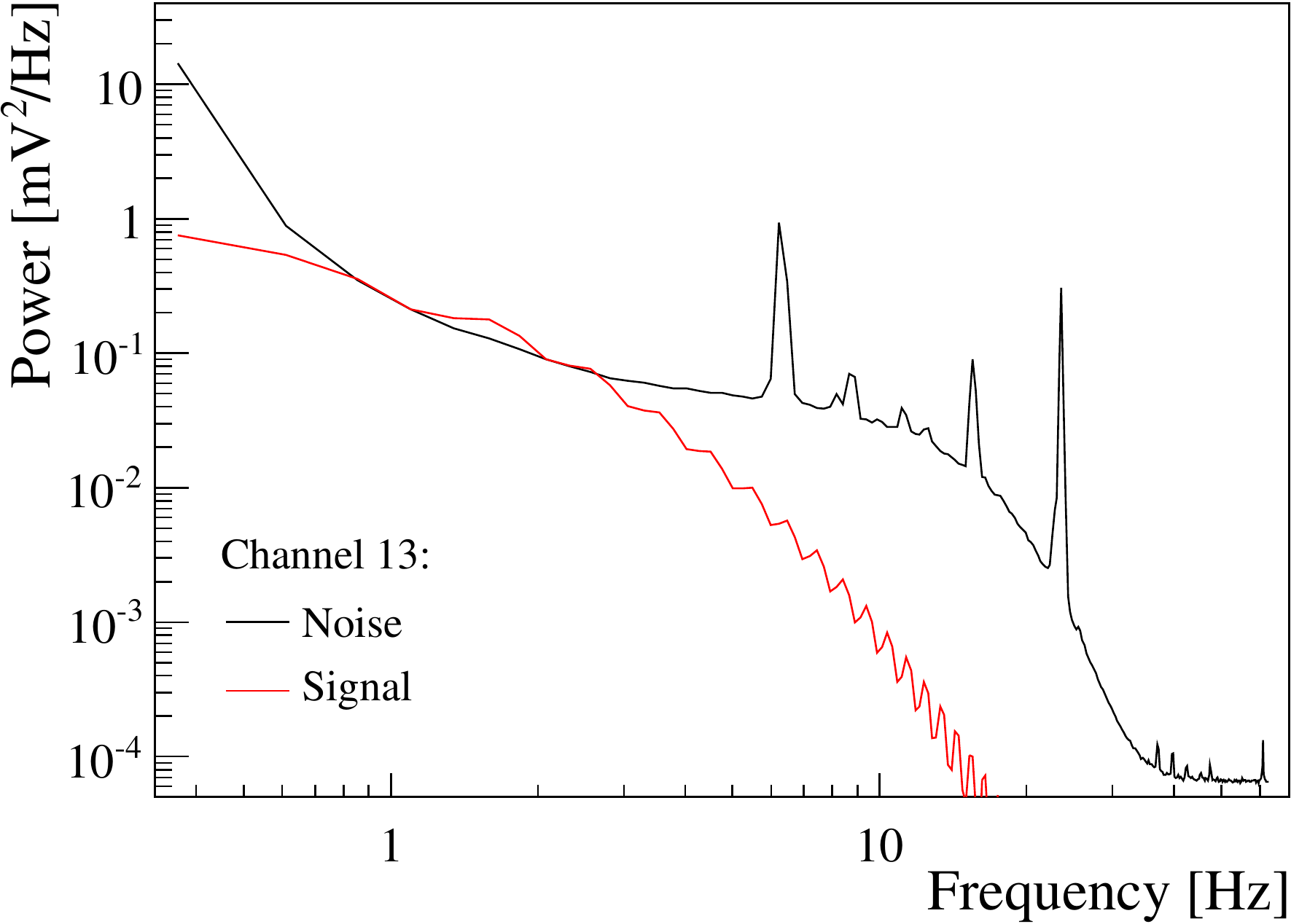}  
\end{minipage}
\caption{Response of a \Cuoricino\ bolometer (channel 13) . Left panel: signal generated by a 1461\un{keV} $\gamma$ particle recorded by the
ADC. Right panel: noise
power spectrum expressed in $\un{mV^2/Hz}$, and energy spectrum of the signal scaled to fit in the noise range.} \label{fig:pulse_noise}
\end{figure}

The typical response of \Cuoricino\ and \Cuore\ bolometers to particles 
impinging on the crystal is of order $100\un{\mu V/MeV}$. The signal
frequency bandwidth is $0-10\un{Hz}$, while the noise components
extend to higher frequencies.  The signal is amplified, filtered with
a 6-pole active Bessel filter with a cut-off frequency of 12\un{Hz}
and then acquired with an 18-bit ADC with a sampling frequency of
125\un{Hz}.  The gain of the amplifiers ranges from 500 to 10000\un{V/V}, 
and is tuned for each bolometer to fit the signals in the ADC range,
which is $\pm10.5\un{V}$.
A typical signal recorded by the ADC, produced
by a 1461\un{keV} $\gamma$ particle fully absorbed in a \Cuoricino\
bolometer, is shown in the left panel of Fig.~\ref{fig:pulse_noise}.

\begin{figure}[tbp]
\centering
\begin{minipage}{0.22\textwidth}
\includegraphics[clip=true,width=1\textwidth]{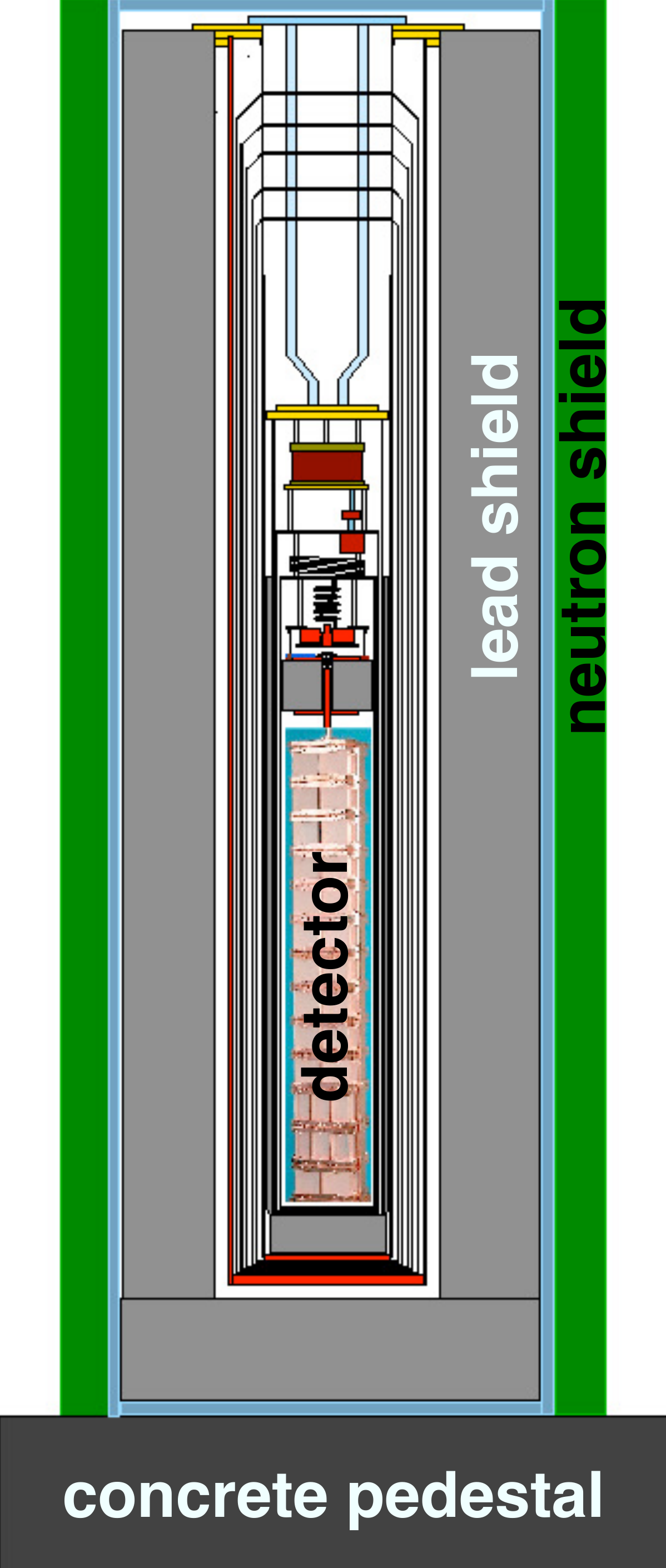}  
\end{minipage}
\hfill
\begin{minipage}{0.6\textwidth}
\includegraphics[clip=true,width=1\textwidth]{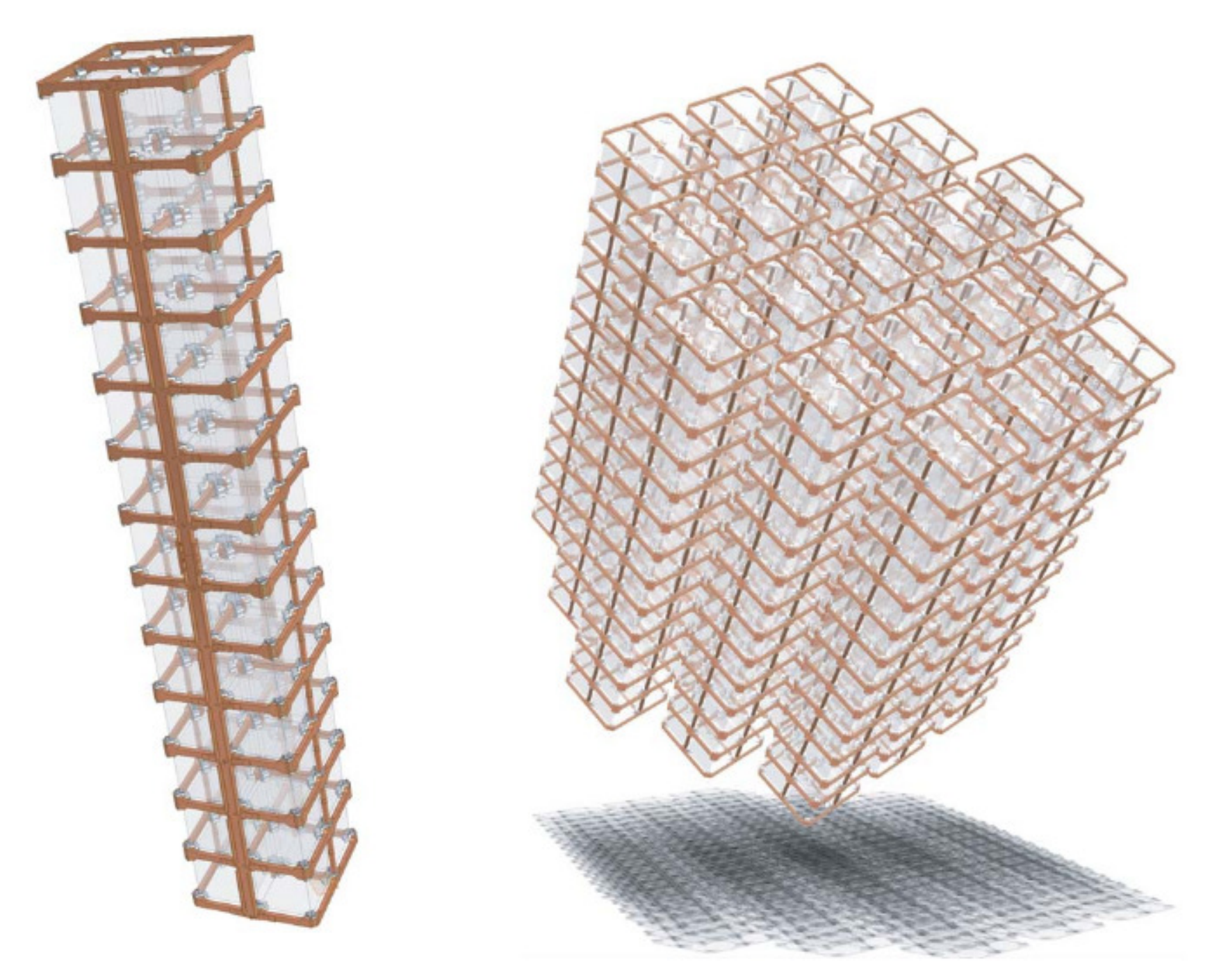}  
\end{minipage}
\caption{Arrays of \TEO\ bolometers. The \Cuoricino\ tower (left), composed
by 62 bolometers, represented inside a schema of the cryostat; a single \Cuore\ tower (middle)
and the entire  \Cuore\ array (right).} \label{fig:cuoricino_and_cuore}
\end{figure}

The 62 bolometers of \Cuoricino\ were arranged in a tower of 13 floors:
11 floors were composed by 4  5x5x5~cm$^{3}$ bolometers each, 2 floors
were composed by 9 small bolometers (3x3x6~cm$^3$) each.  The small
bolometers were recycled from previous arrays, and will not be used
in \Cuore, which will only use 5x5x5~cm$^{3}$ crystals arranged in
19 towers of 13 floors each (Fig.~\ref{fig:cuoricino_and_cuore}).
The front-end electronics, which provide the bias, the load resistors
and the amplifier, are placed outside the cryostat, at room temperature~\cite{AProgFE}. 
In \Cuoricino, 1/3 of the electronics channels used load resistors and pre-amplifiers 
located inside the cryostat, at 110\un{K}. This was done to reduce the 
noise from load resistors, but since no improvement was achieved, the cold
electronics will not be used in \Cuore.
Three channels (1 cold, 2 warm) were lost after the initial \Cuoricino\ cool-down, reducing the number of active channels
to 59.

Vibrations of the detector structure generate two types of noise: thermal
and microphonic. The thermal noise is due to vibrations of the crystals
that induce temperature fluctuations of the crystals themselves. The
microphonic noise is due to vibrations of the wires that connect the thermistor
to the cryostat socket. A typical noise power spectrum is
shown in the right panel of Fig.~\ref{fig:pulse_noise}; the peaks in the
figure are attributed to microphonism, while the continuum is attributed
to crystal vibrations.

The data acquision system connected to the ADC boards implements two kind of software
triggers. The first trigger fires when a
signal is detected on a bolometer, the second fires randomly in time
to acquire noise waveforms. To estimate the noise correlation, the
acquisition was programmed to acquire simultaneously all the bolometers of
\Cuoricino\ when the signal or the noise triggers fired on a bolometer.
The length of the acquisition window was set to $M=512$ samples on all channels,
corresponding to 4.096\un{s}.
The data used in this paper amount to 2 days taken from the last month of \Cuoricino\ operation.

\section{Noise correlation} 

The noise power spectrum is estimated from a large number of waveforms
acquired with the random trigger, removing those
that, by chance, contain signals.  The power spectrum $N_i(\omega)$
of each bolometer channel $i$ is computed as:
\begin{equation}
N_i(\omega) =  <n_i(\omega) \cdot n_i^*(\omega)>\,,
\label{eq:nps}
\end{equation}
where  $<>$ denotes the average taken
over a large number of waveforms and $n(\omega)$ is the discrete Fourier transform
(DFT) of a noise waveform $n_i(t)$.  To avoid DFT artifacts, each waveform
is weighted before the DFT using a Welch windowing function:
\begin{equation}
n_i(t) \rightarrow n_i(t) \cdot w(t)\;,\quad
w(t) = \frac{1}{G}\left\{ \frac{M}{2} - \left[\left(t-\frac{M}{2}\right)\right]^2\right\}\;.
\end{equation}
In the above expression the parameter $G$ is a normalization coefficient
chosen to satisfy the condition~\cite{Novotny2008236}:
\begin{equation}
\sum_{\omega} N_i(\omega) = \sigma_i^2
\end{equation}
where $\sigma_i$ is the standard deviation of the noise in the time domain.

\begin{figure}
\centering
\includegraphics[clip=true,width=0.8\textwidth]{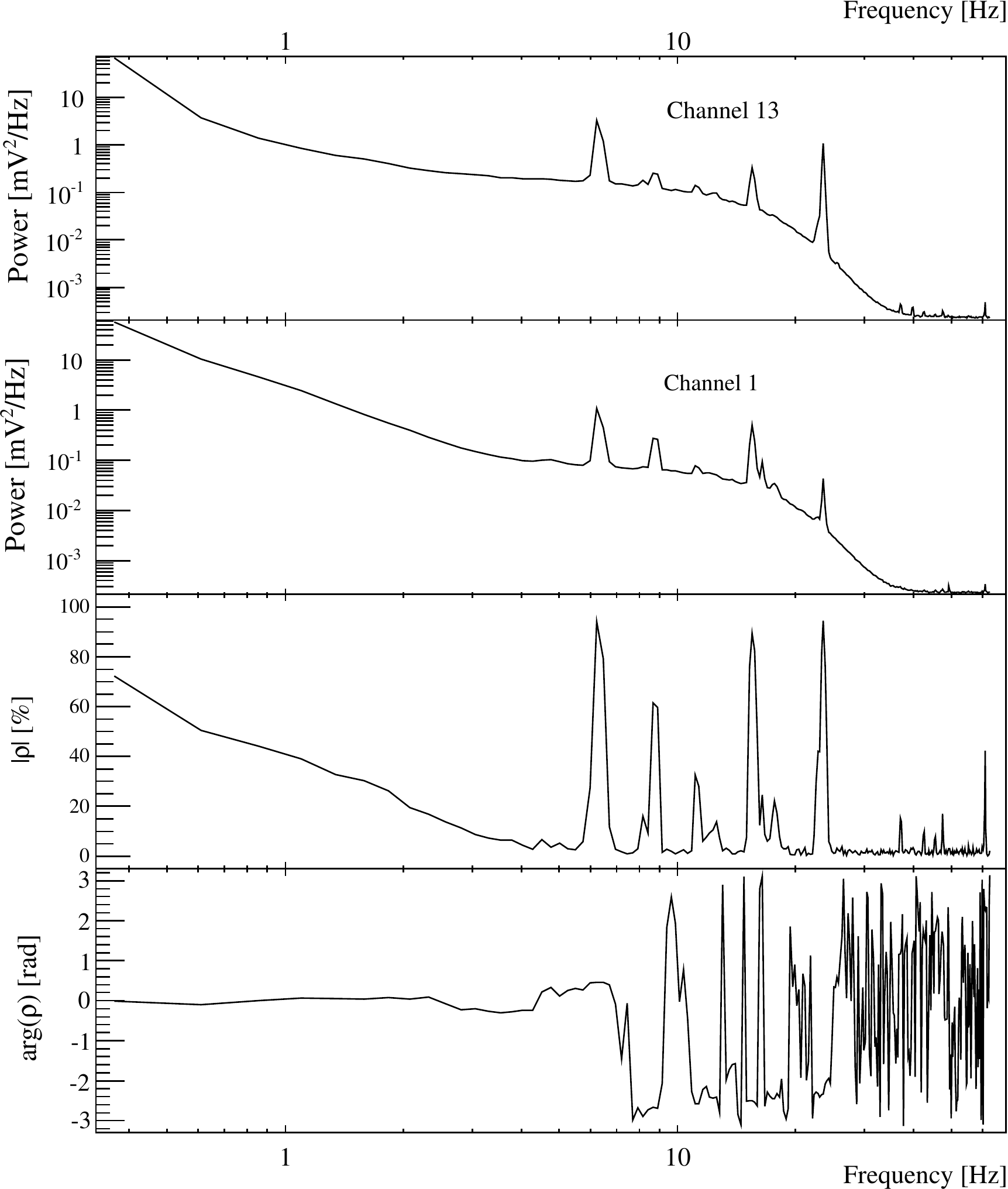}  
\caption{Power spectra of \Cuoricino\ channels 13 and 1 and their correlation $\rho(\omega)$.}
\label{fig:correlation_131}
\end{figure}

The covariance between the noises on bolometers channels $i$ and $j$ is estimated
as:
\begin{equation}
N_{ij}(\omega) = <n_i(\omega) \cdot n_j^*(\omega)>
\label{eq:cov}
\end{equation}
where the definitions are the same of Eq.~\ref{eq:nps}.  The magnitude
of $N_{ij}(\omega)$ is the covariance as usually intended for real variables,
while the phase accounts for any possible time delay of the frequency $\omega$
observed in the two bolometers.  The correlation is also a complex
quantity, and is defined as:
\begin{equation}
\rho_{ij}(\omega)  = \frac{{\rm N}_{ij}(\omega)}{\sqrt{{N}_i(\omega) {N}_j(\omega)}}\quad.
\end{equation}
As an example, we show the behaviour of  $\rho_{13,1}(\omega)$ in Fig.~\ref{fig:correlation_131}; the microfonic peaks and the low frequencies are highly correlated,
while the intermediate frequencies are almost uncorrelated.
The pattern in the figure is visible for many $(i,j)$ pairs and indicates that the source of correlated noise has analogous origin in all bolometers.

\begin{figure}
\centering
\includegraphics[clip=true,width=1\textwidth]{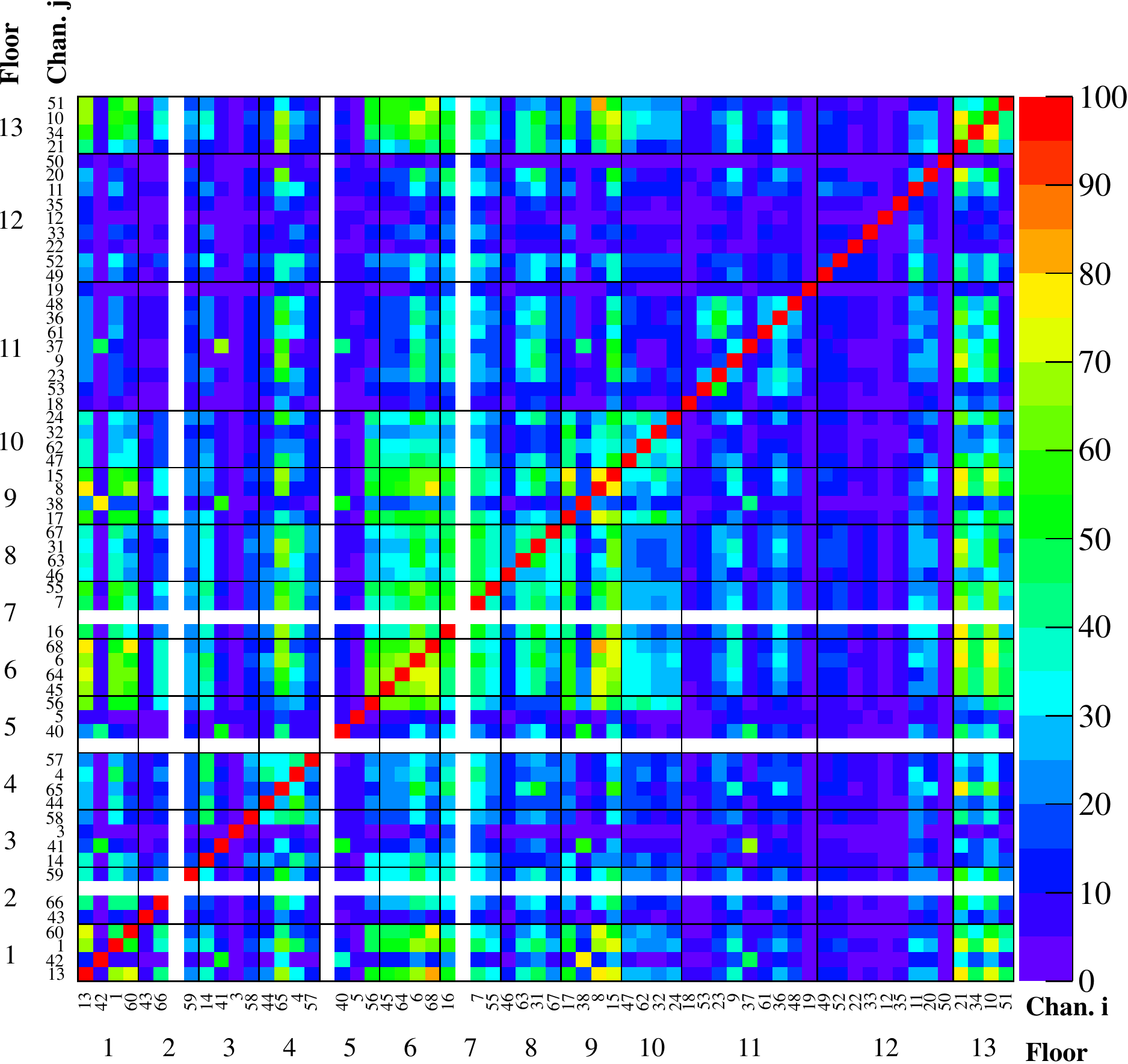}  
\caption{Noise correlation in \Cuoricino. Channels are ordered as bolometers are ordered in the array. White lines indicate dead channels.}
\label{fig:avgcorr_space}
\end{figure}

The average correlation over all frequencies can be computed as:
\begin{equation}
\overline{\rho_{ij}} = \sqrt{
\frac{\sum_\omega |\rho_{ij}(\omega)|^2 N_i(\omega)}
{\sum_\omega  N_i(\omega)}
}\;,
\label{eq:avgcorr}
\end{equation}
and is found to be around 60\% for the $(i,j)$ pair $(13,1)$. The average correlation matrix, including the correlation of all the \Cuoricino\ channels,
is shown in Fig.~\ref{fig:avgcorr_space}. The matrix is not exactly triangular because Eq.~\ref{eq:avgcorr} is not symmetric under the exchange of  channels $i$ and $j$,  since the same frequency
may contribute with a different weight to $N_i(\omega)$ and $N_j(\omega)$. 
The figure shows that the correlation is high,  even tough there is no evident link with the spatial disposition of the bolometers. 
Bolometers close to each other, in fact, are not necessarily the most correlated. 

We also tried to order the channels by their cabling group in the cryostat or by cold/warm electronics channels, but also in this case we did not see any link.
Other variables linked to the noise correlation could be, for example, the path of the wires along the tower, the tightness of the crystals and the length of thermistor wires. These variables are not accessible now since \Cuoricino\ has been disassembled. Moreover, while the understanding 
of the noise sources is important, in this paper we focus on the analysis algorithms to reduce the observed noise.
 
\section{Noise decorrelation}

In \TEO\ bolometers the noise is purely additive to the signal, meaning
that the waveform $f(t)$ observed in presence of a
signal of amplitude $A$ can be expressed as:
\begin{equation}
f(t) = A\cdot s(t) + n(t)\;,
\label{eq:signal+noise}
\end{equation}
where $s(t)$ is the signal shape.
Since the noise on different bolometers is partially
correlated, it is possible to remove part of $n(t)$ from $f(t)$
using the waveforms from other bolometers in which only noise is present.
In the following we develop a method to remove the correlated noise.

Assuming that each frequency component of $n_i(\omega)$ is normally
distributed, the multidimensional probability distribution of the noises
on all bolometers $\vec{n}(\omega) = \{n_1(\omega),n_2(\omega),\ldots\}$ can be written as:
\begin{equation}
P[\vec{n}(\omega)] \propto  \exp \left( -\frac{1}{2} \vec{n}^{\dag}(\omega) \hat{N}^{-1}(\omega) \vec{n}(\omega) \right)
\label{eq:gaussianamultidim}
\end{equation}
where $\hat{N}$ is the covariance matrix whose elements are defined
in Eq.~\ref{eq:cov}.  The probability distribution of  the noise on
bolometer  $i$, $n_i(\omega)$, can be obtained integrating out the noise of other bolometers:
\begin{equation}
P[n_i(\omega)]\propto \exp \left[-\frac{N^{-1}_{ii}(\omega)}{2}  \left(n_{i}(\omega) + \sum_{j \neq i} \frac{N_{ij}^{-1}(\omega)}{N_{ii}^{-1}(\omega)} n_{j}(\omega) \right)^{2} \right]  \;.
\end{equation}
The above equation represents a Gaussian with mean 
$\left[-\sum_{j \neq i} \frac{N_{ij}^{-1}(\omega)}{N_{ii}^{-1}(\omega)} n_{j}(\omega)\right]$
and variance
$1/{N_{ii}^{-1}(\omega)}$.
The decorrelated value of $n_{i}(\omega)$ can be
obtained for each waveform as:
\begin{equation}
n^d_{i}(\omega)=n_{i}(\omega)+\sum_{j \neq i} \frac{N_{ij}^{-1}(\omega)}{N_{ii}^{-1}(\omega)} n_{j}(\omega) \;,
\label{eq:noise_decorrelated}
\end{equation}
and its power spectrum is expected to be:
\begin{equation}
N^d_i(\omega) = {\frac{1}{N_{ii}^{-1}(\omega)}}\;.
\label{eq:nps_decorrelated}
\end{equation}
A generic waveform $f_i(t)$ on bolometer $i$, containing noise or noise+signal,
can be decorrelated using the noise from all other bolometers as:
\begin{equation}
f^d_{i} (\omega)= f_{i}(\omega) +\sum_{j \neq i} \frac{N_{ij}^{-1}(\omega)}{N_{ii}^{-1}(\omega)} n_{j}(\omega)
= A\, s_i(\omega) + n^d_i(\omega)\;.
\label{eq:signal+noise_decorrelated}
\end{equation}

Summarizing, once the covariance matrix in Eq.~\ref{eq:cov} is estimated from the data, the 
waveforms on each bolometer are decorrelated using Eq.~\ref{eq:signal+noise_decorrelated} 
and the effect on the noise power spectrum can be predicted with Eq.~\ref{eq:nps_decorrelated}.

\section{Application to data} 
\begin{figure}[htbp]
\begin{center}
\includegraphics[clip=true,width=0.48\textwidth]{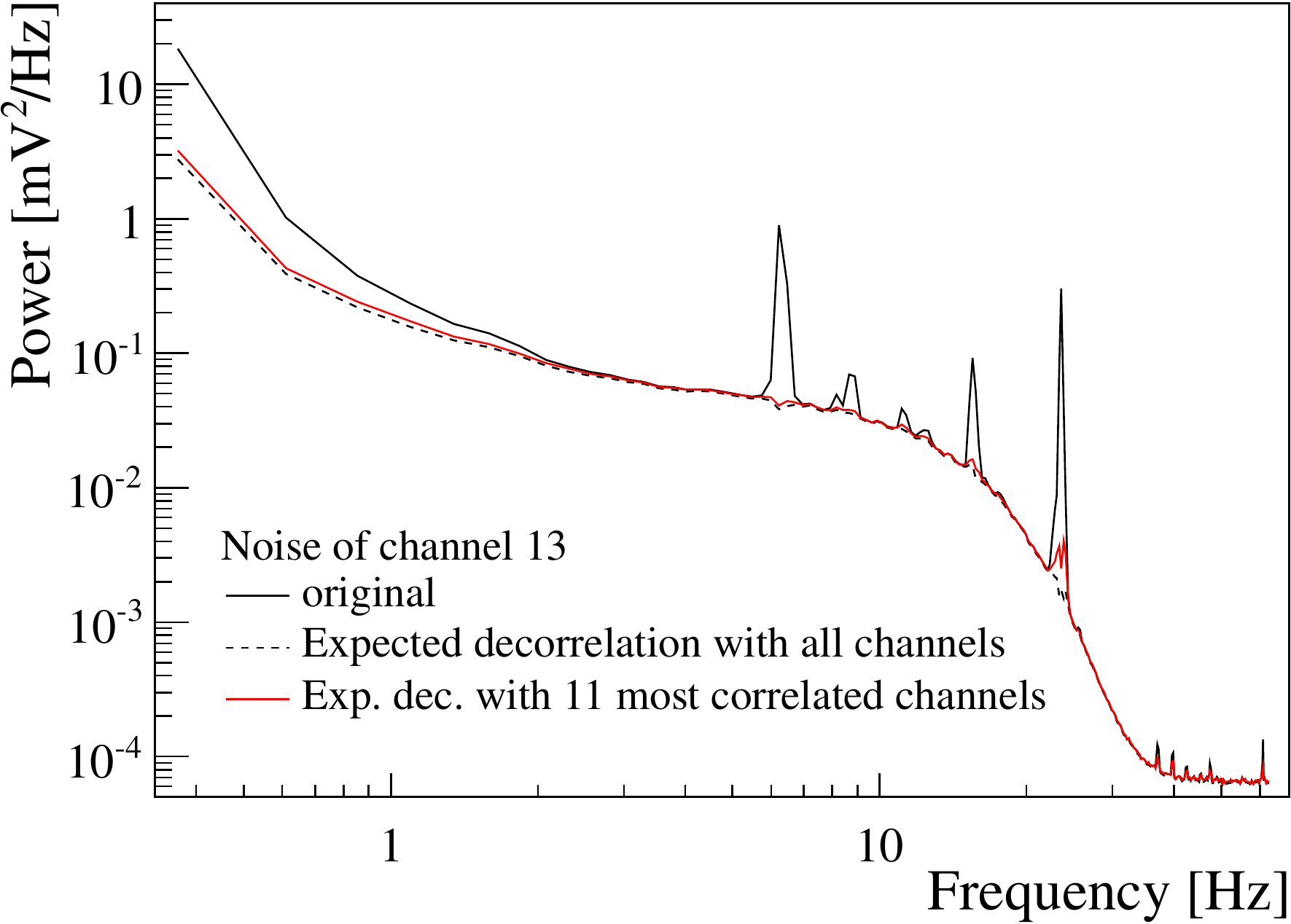}  
\includegraphics[clip=true,width=0.48\textwidth]{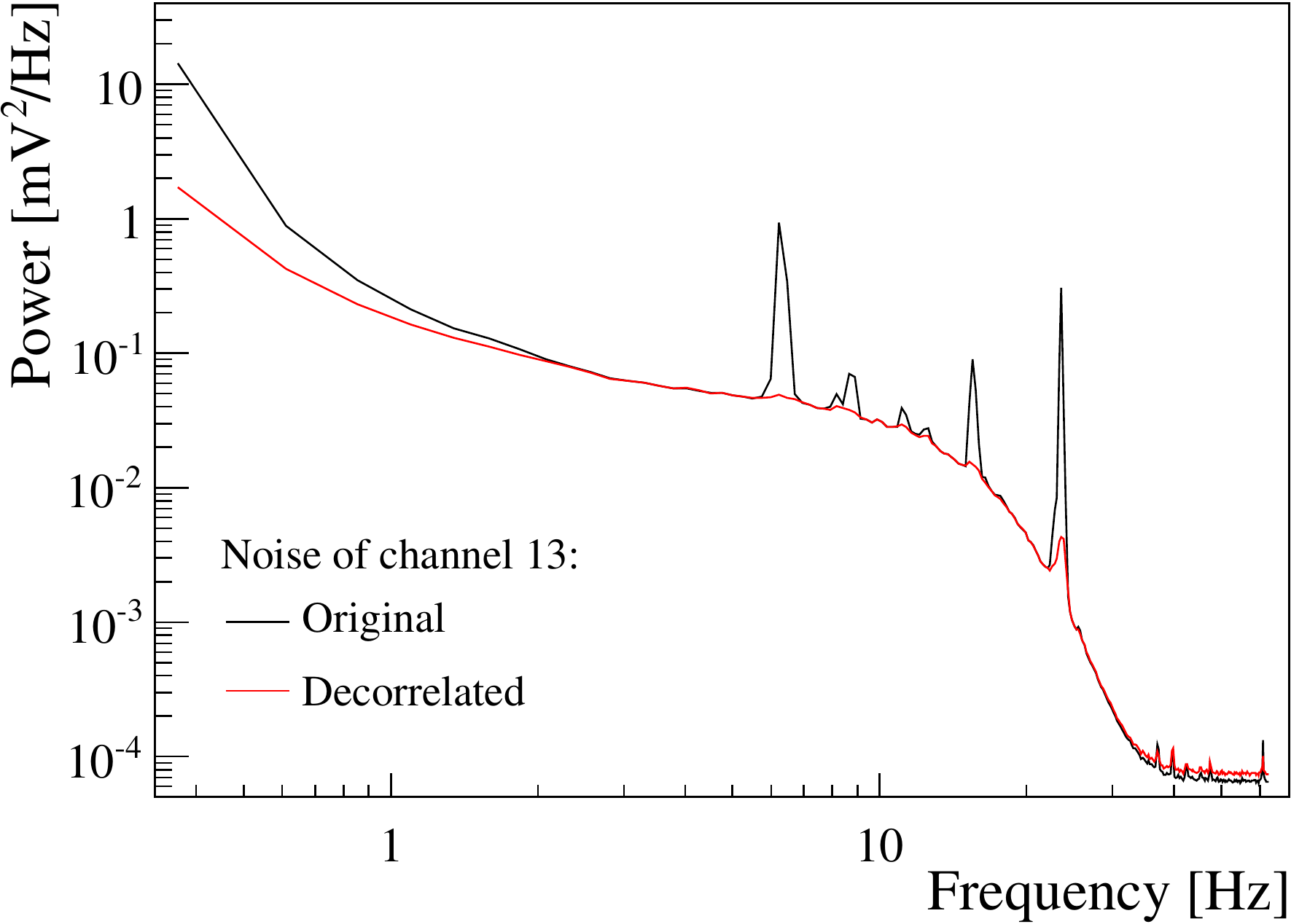}  
\caption{Left: Noise power spectra of bolometer 13: original (solid black),
expected decorrelated using 61 bolometers (dashed black), and expected
decorrelated using the 11 most correlated bolometers (solid red). Right: original noise
power spectrum and power spectrum of waveforms decorrelated using  Eq.~\protect\ref{eq:signal+noise_decorrelated}.
The original power spectra in the figures differs slightly because different sets of data has been used
to estimate the covariance matrix (left) and  the effect of the decorrelation (right).}
\label{fig:potenzaattesatuttoilcucuzzaro}
\end{center}
\end{figure}

The acquired data are split in two sets, one set is used to estimate the correlation matrix and the expected
noise power spectra, the other is used to verify that the application of the decorrelation algorithm to waveforms 
produces results consistent with the expectations. 
Figure \ref{fig:potenzaattesatuttoilcucuzzaro} (left) shows  the original
power spectrum of  channel 13 (Eq.~\ref{eq:nps})
and the one expected decorrelating from
all the other \Cuoricino\ bolometers (Eq.~\ref{eq:nps_decorrelated}).  The power of low frequencies is reduced
and the microphonic peaks are completely removed. 
In principle one could be satisfied and proceed to apply
Eq.~\ref{eq:signal+noise_decorrelated} to the single waveforms.
However, using all the  bolometers to decorrelate every triggered waveform
is expensive from the computational point of view, since it requires
 DFTs for each bolometer used.  Moreover all the waveforms used to
decorrelate should not contain signals. This requirement is often
fulfilled in \Cuoricino, where the overall counting rate was around 0.1\un{Hz}, but not in \Cuore, where
the rate is expected to be larger. For this
reason we tested to see if the decorrelation is effective using a smaller number 
of bolometers. For each bolometer we selected the  most correlated
bolometers and we computed the expected decorrelated power spectrum. The results obtained are equivalent to those
obtained using all bolometers (Fig.~\ref{fig:potenzaattesatuttoilcucuzzaro} left). The number of bolometers used to
decorrelate was set to 11, a number sufficiently high to maximize the decorrelation and sufficiently low to
ensure good performances.

\begin{figure}[htbp]
\begin{center}
\includegraphics[clip=true,width=0.48\textwidth]{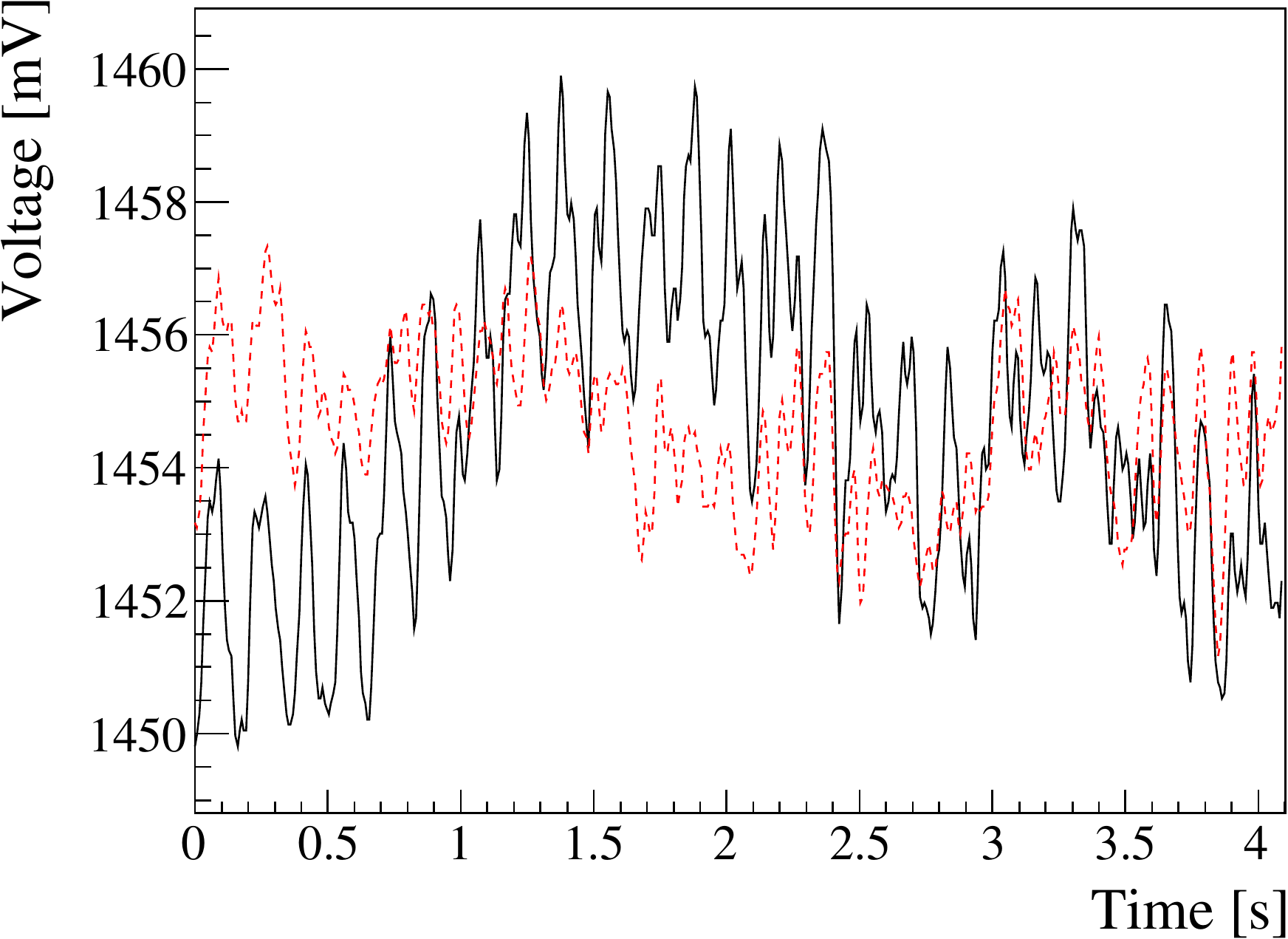}  
\hfill
\includegraphics[clip=true,width=0.48\textwidth]{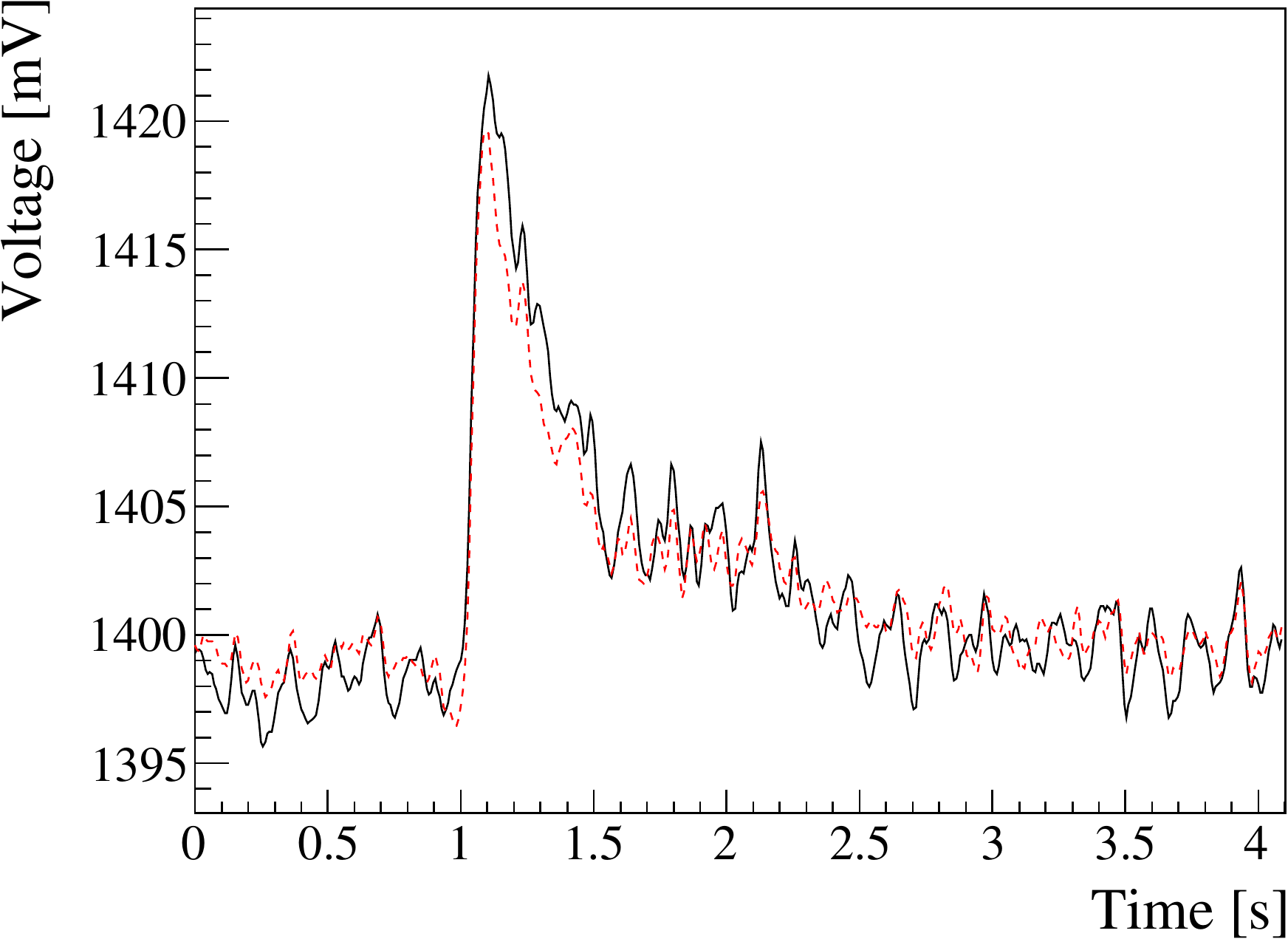}  
\caption{Left: original (solid black) and decorrelated (dashed red) noise waveform from channel 13. Right: signal generated by a 43 keV $\gamma$ particle.
The decorrelation algorithm reduces the noise leaving the signal unmodified.} \label{fig:decowave}
\end{center}
\end{figure}

\begin{figure}[htbp]
\begin{center}
\includegraphics[clip=true,width=0.95\textwidth]{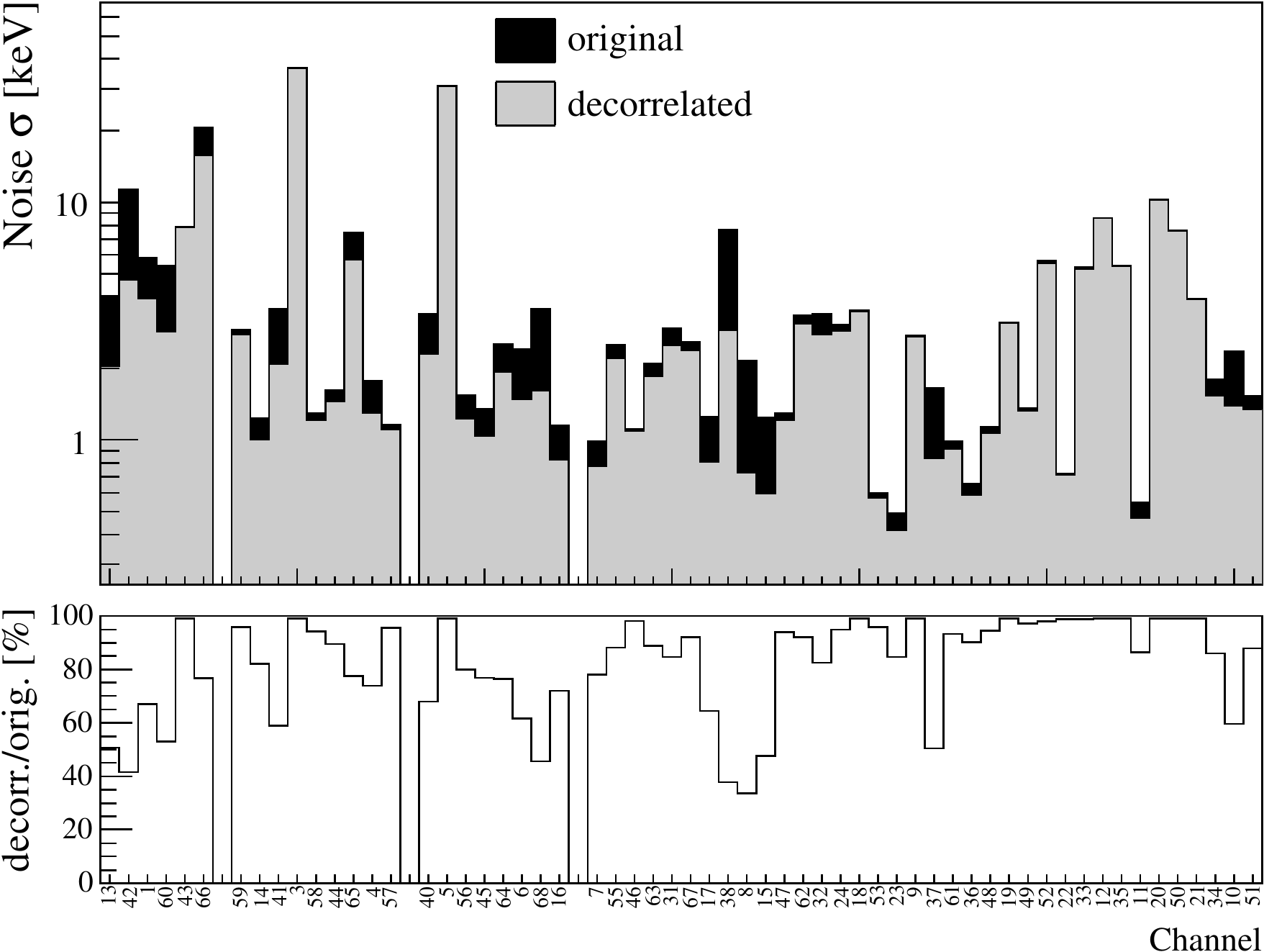}
\caption{Noise resolution of all \Cuoricino\ bolometers before
and after the decorrelation. Values are expressed in keV.} \label{fig:rms}
\end{center}
\end{figure}

The noise power spectrum obtained on waveforms decorrelated with
Eq.~\ref{eq:signal+noise_decorrelated} is shown in
Fig.~\ref{fig:potenzaattesatuttoilcucuzzaro} (right).  
The decorrelation is found to be very effective on channel 13: the noise $\sigma$ is reduced from 2.0\un{mV} to 1.2\un{mV}
(corresponding to 3.6 and 2.2~keV, respectively),
an effect which is  visible also in the time domain (Fig.~\ref{fig:decowave}).
We show in Fig.~\ref{fig:rms} the energy resolution on all
bolometers before and after the decorrelation: in a few bolometers the resolution remains the same, while in others
it is reduced up to 50\%.

\section{Combination with the optimum filter} 

In the data analysis of \TEO\ bolometers the signal amplitude is estimated
using the optimum filter~\cite{Gatti:1986cw,Radeka:1966}, a filter that
significantly improves the energy resolution. This filter can be used when the shape
of the signal and the noise power spectrum of each channel are known.
The filter acts on a single channel and its transfer function is
\begin{equation}
H(\omega) =  h \frac{s^*(\omega)}{N(\omega)}e^{-\jmath\, \omega i_M }\,,
\label{eq:of}
\end{equation}
where $i_M$ is a parameter to adjust the delay of the
filter. In this work, as in Ref.~\cite{Didomizio:2010ph}, we chose it equal
to the maximum position of $s_i$, so that the maximum of the filtered
signal is aligned to the maximum of the non-filtered one. $h$ is a normalization
constant that leaves unmodified the amplitude of the signal:
$h = \left[\sum_\omega \frac{|s(\omega)|^2}{N(\omega)}\right]^{-1}$~.

\begin{figure}[htbp]
\begin{center}
\includegraphics[clip=true,width=0.48\textwidth]{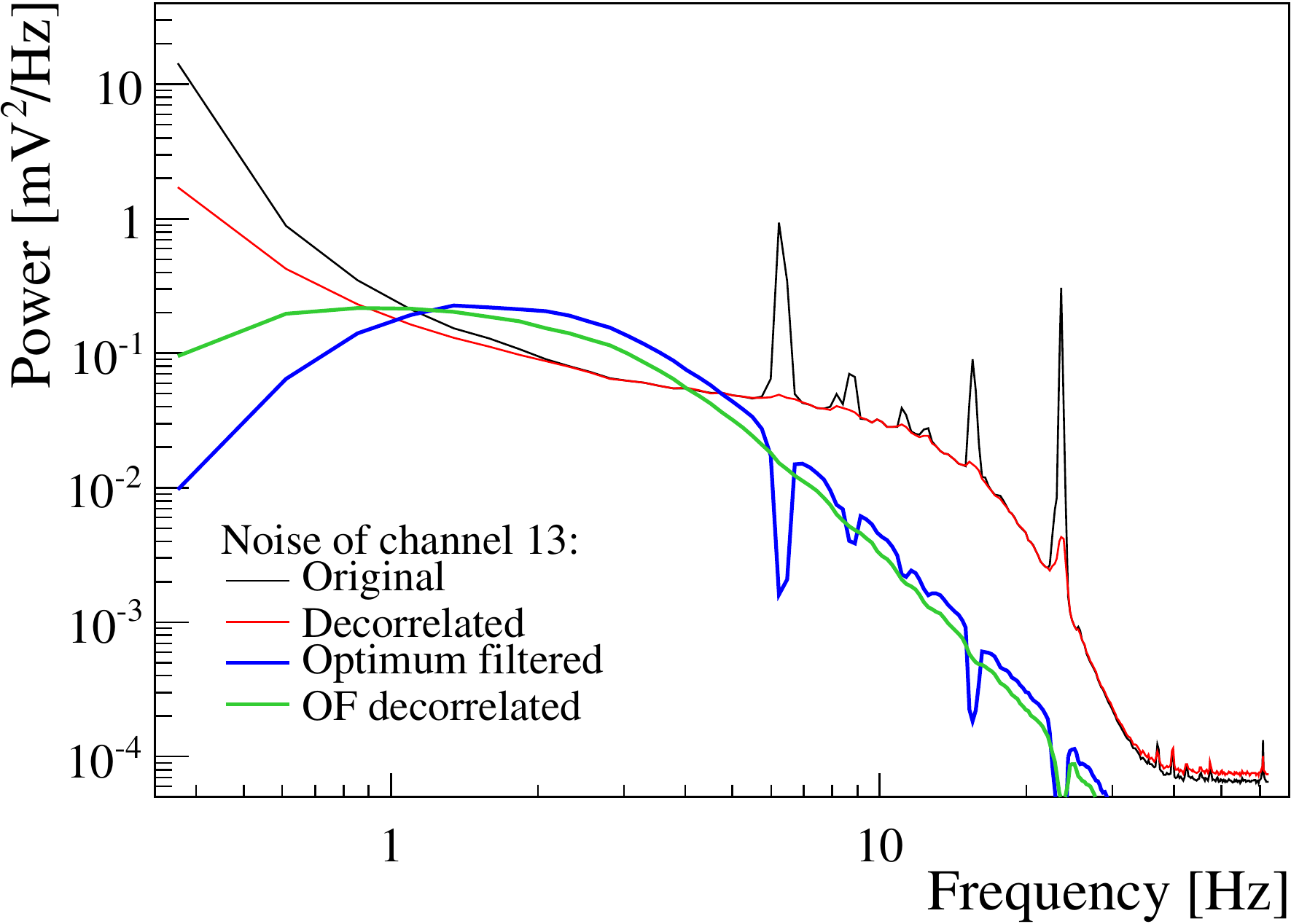}  
\includegraphics[clip=true,width=0.48\textwidth]{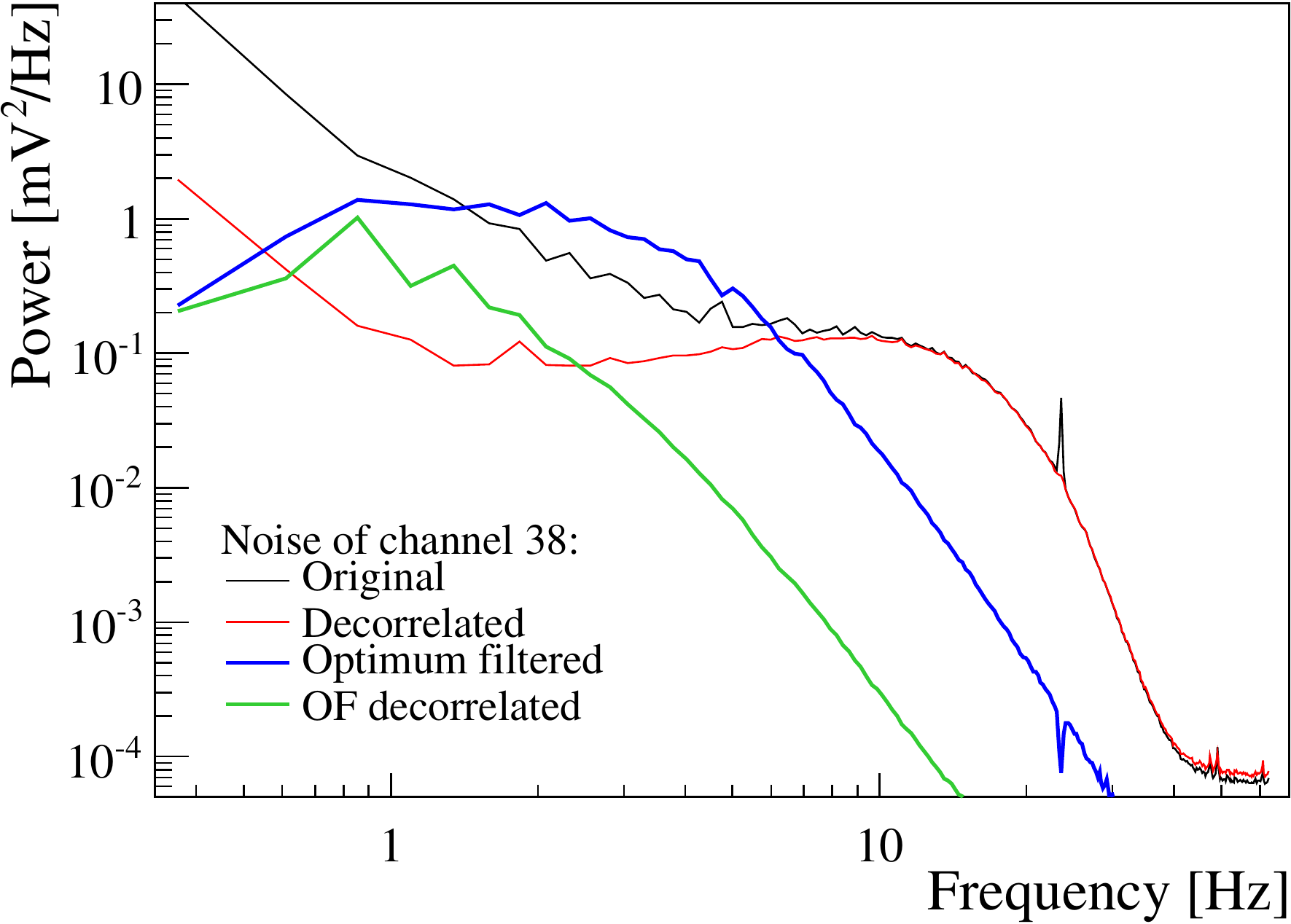}  
\caption{Comparison of original, decorrelated, optimum filtered and decorrelated/optimum filtered noises
on bolometers 13 (left) and 38 (right).}
\label{fig:of_filtered}
\end{center}
\end{figure}

To combine the optimum filter with the decorrelation, it is sufficient to
replace in Eq.~\ref{eq:of} the original power spectrum $N(\omega)$  with
$N^d(\omega)$ defined in Eq.~\ref{eq:nps_decorrelated}, and then process
the decorrelated waveforms in Eq.~\ref{eq:signal+noise_decorrelated}. The
comparison of the filtered noise power spectrum with and without the
decorrelation for two bolometers with opposite behavior (channels 13 and 38)
is shown in Fig.~\ref{fig:of_filtered}. In channel 13, while the
decorrelated noise is lower than the original one, there is no big improvement
in the filtered-decorrelated noise, because the correlation is high only on frequencies
with low signal to noise ratio. In channel 38, on the other hand,
the decorrelation lowers the noise after the optimum filter.

\begin{figure}[htbp]
\begin{center}
\includegraphics[clip=true,width=0.95\textwidth]{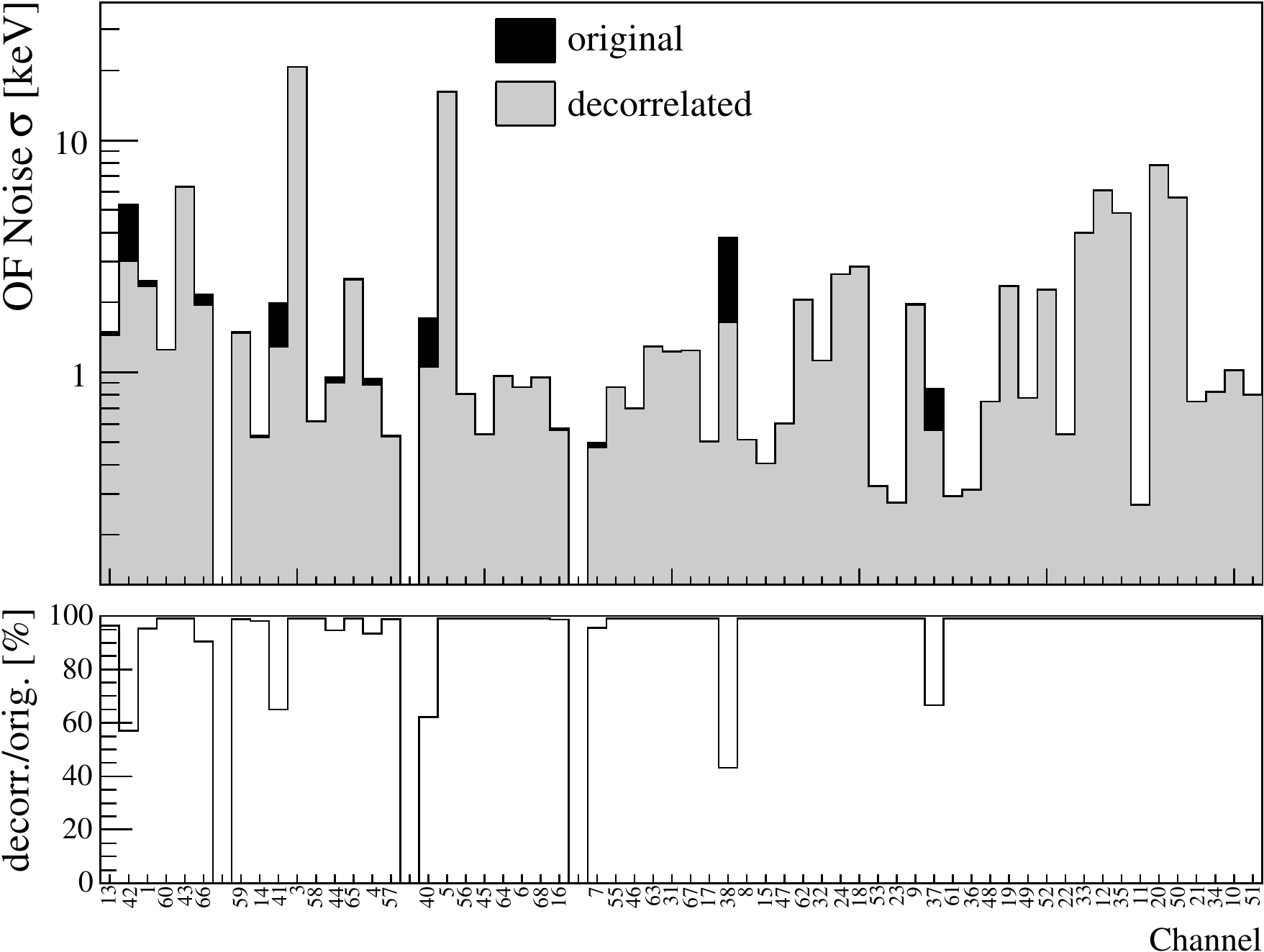}
\caption{Noise resolution  of all \Cuoricino\
bolometers after the optimum filter before and after  the decorrelation.}
\label{fig:rmsof}
\end{center}
\end{figure}

To summarize we compare the expected resolution of the entire array
using both decorrelation and optimum filter with respect to optimum
filter only (Fig.~\ref{fig:rmsof}). The improvement is not significant
as when we decorrelated the original waveforms (Fig.
~\ref{fig:rms}), because in most cases the noise is not correlated in the signal frequency bandwidth.

\section{Conclusions} 

In this paper we developed a method to remove the correlated noise
between different detectors. The application to \Cuoricino\ showed that the correlated noise, 
generated by vibrations of the detector structure, can be efficiently removed.  However, when the
decorrelation is combined with the optimum filter, the resolution does
not improve significantly since the correlated noise lies at frequencies higher than the
signal frequency bandwidth.  \Cuore, a \TEO\ bolometric array 20 times
larger than \Cuoricino, will have a new detector structure that could
induce  different vibrational noise. Moreover, while \Cuoricino\ was refrigerated using
liquid helium, the \Cuore\ cryostat will use pulse tubes, acoustic cryocoolers oscillating at
low frequencies (around 2 Hz). If the noise will be
unfortunately large in the signal band, the method we introduced will be a
valid tool to improve the \Cuore\ performances.

\acknowledgments
We are very grateful to the members of the \Cuoricino\ and \Cuore\ collaborations, in particular to 
F.~Bellini, L.~Cardani, T.~Gutierrez, K.~Han and G.~Pessina for their precious suggestions.
Special thanks go to Robert Joachim, for the preliminary studies he did during his summer student fellowship at INFN Roma.
\bibliographystyle{JHEP} 
\bibliography{main}

\providecommand{\href}[2]{#2}\begingroup\raggedright\begin{thebibliography}{10}

\bibitem{Ardito:2005ar}
R.~Ardito {\em et.~al.}, {\it {CUORE: A Cryogenic Underground Observatory for
  Rare Events}},  \href{http://xxx.lanl.gov/abs/hep-ex/0501010}{{\tt
  hep-ex/0501010}}.

\bibitem{ACryo}
C.~Arnaboldi {\em et.~al.}, {\it {CUORE}: {A} {C}ryogenic {U}nderground
  {O}bservatory for {R}are {E}vents},  {\em Nucl. Instrum. Meth. A.} {\bf 518}
  (2004) 775, [\href{http://xxx.lanl.gov/abs/hep-ex/0212053v1}{{\tt
  hep-ex/0212053v1}}].

\bibitem{Didomizio:2010ph}
S.~Di~Domizio, F.~Orio, and M.~Vignati, {\it {Lowering the energy threshold of
  large-mass bolometric detectors}},  {\em JINST} {\bf 6} (2011) P02007,
  [\href{http://xxx.lanl.gov/abs/1012.1263}{{\tt arXiv:1012.1263}}].

\bibitem{Andreotti:2010vj}
E.~Andreotti {\em et.~al.}, {\it {$^{130}$Te Neutrinoless Double-Beta Decay
  with CUORICINO}},  {\em Astropart.Phys.} {\bf 34} (2011) 822,
  [\href{http://xxx.lanl.gov/abs/1012.3266}{{\tt arXiv:1012.3266}}].

\bibitem{Bellini:2010iw}
F.~Bellini {\em et.~al.}, {\it {Response of a $TeO_{2}$ bolometer to alpha
  particles}},  {\em JINST} {\bf 5} (2010) P12005,
  [\href{http://xxx.lanl.gov/abs/1010.2618}{{\tt arXiv:1010.2618}}].

\bibitem{wang}
N.~Wang {\em et.~al.}, {\it Electrical and thermal properties of
  neutron-transmutation-doped {G}e at 20 m{K}},  {\em Phys. Rev. B} {\bf 41}
  (1990) 3761.

\bibitem{Itoh}
K.~M. {Itoh} {\em et.~al.}, {\it Neutron transmutation doping of isotopically
  engineered {G}e},  {\em {A}ppl. {P}hys. {L}ett.} {\bf 64} (1994) 2121.

\bibitem{AProgFE}
C.~Arnaboldi {\em et.~al.}, {\it The programmable front-end system for
  {CUORICINO}, an array of large-mass bolometers},  {\em IEEE Trans. Nucl.
  Sci.} {\bf 49} (2002) 2440.

\bibitem{Novotny2008236}
M.~Novotny and M.~Sedlacek, {\it {RMS} value measurement based on classical and
  modified digital signal processing algorithms},  {\em Measurement} {\bf 41}
  (2008) 236.

\bibitem{Gatti:1986cw}
E.~Gatti and P.~F. Manfredi, {\it Processing the signals from solid state
  detectors in elementary particle physics},  {\em Riv. Nuovo Cimento} {\bf 9}
  (1986) 1.

\bibitem{Radeka:1966}
V.~Radeka and N.~Karlovac, {\it Least-square-error amplitude measurement of
  pulse signals in presence of noise},  {\em Nucl. Instrum. Methods} {\bf 52}
  (1967) 86.

\end{thebibliography}\endgroup

\end{document}